\documentclass[10pt,a4paper]{article}
\usepackage{amsmath,amssymb}
\begin{document}
\title{Two Classes of Crooked Multinomials Inequivalent to Power Functions}
\author{Xueying Duan\footnote{Department of Electronic Engineering, Hunan University of Science and Engineering,
Yongzhou, China,  425100 \ E-mail: qcwang@fudan.edu.cn.} \and
Qichun Wang\footnote{Department
of mathematics, Hunan
University of Science and Engineering, Yongzhou, China,  425100 \
E-mail: duanduan169@163.com.}} 
\date{}
\maketitle
\begin{abstract}
%\boldmath
It is known that crooked functions can be used to construct many
interesting combinatorial objects, and a quadratic function is
crooked if and only if it is almost perfect nonlinear (APN). In this
paper, we introduce two infinite classes of quadratic crooked
multinomials on fields of order $2^{2m}$. One class of APN functions
constructed in [7] is a particular case of the one we construct in
Theorem 1. Moreover, we prove that the two classes of crooked
functions constructed in this paper are EA inequivalent to power
functions and conjecture that CCZ inequivalence between them also
holds.
\end{abstract}

%\par {\bf MSC 2010:} 11T71

\section{Introduction}
Let $F_{2^n}$ denote a finite field with $2^n$ elements, which is
also considered as an $n$-dimensional vector space over its subfield
$F_{2}$. The affine hyperplanes in $F_{2^n}$ are the subspaces of
dimension $n-1$ and their complements. A function
$f:F_{2^n}\rightarrow F_{2^n}$ is called differentially
$\delta$-uniform [26] if for every $a\neq 0$ and every $b$ in
$F_{2^n}$, the equation $f(x)+f(x+a)=b$ has at most $\delta$
solutions. Vectorial Boolean functions used as S-boxes in block
ciphers should have low differentially uniformity to resist
differential cryptanalysis [6]. Since for any function, we have
$\delta\geq 2$ (if $t$ is a solution, then $t+a$ is a solution too),
differentially 2-uniform functions, called almost perfect nonlinear
(APN), are optimal.
\par For odd $n$, the property of APN is closely related to another
extremal kind of nonlinearity, called almost Bent (AB), which can be
described by the walsh transform. Let $tr(x)$ denote the trace
function from $F_{2^n}$ to $F_2$ (that is,
$tr(x)=x+x^2+...+x^{2^{n-1}}$), then the Walsh transform of a given
function $f: F_{2^n} \rightarrow F_{2}$ is the integer-valued
function over $F_{2^n}$ which is defined as
\[
W_f(\omega)=\sum_{x\in F_{2^n}} (-1)^{f(x)+tr(\omega x)}.
\]
The nonlinearity of $f$ is given by
\[
nl(f)=2^{n-1}-\frac{1}{2}\max_{\omega \in F_{2^n}}|W_f(\omega)|.
\]
The Walsh transform of $f: F_{2^n} \rightarrow F_{2^n}$ is defined
as the collection of all Walsh transforms of component functions of
$f$, i.e.,
\[
W_f(\omega,a)=\sum_{x\in F_{2^n}} (-1)^{tr(af(x))+tr(\omega x)}.
\]
The set
\[
\Gamma_f=\{W_f(\omega,a): \omega,a\in F_{2^n}, a\neq 0\}
\]
is called the Walsh spectrum of $f$. If the Walsh spectrum of $f$
equals $\{0,\pm2^{\frac{n+1}{2}}\}$, then the function $f$ is called
AB [15]. Every AB function is APN, and for $n$ odd, any quadratic
function is APN if and only if it is AB. There are many papers on
these two notions (see [2,3,9,11-14,16-19,21,22]).
\par The APN function $f(x)$ is called crooked if the set $\{f(x)+f(x+a): x\in F_{2^n}\}$ is
an affine hyperplane of $F_{2^n}$ for each $0\neq a\in F_{2^n}$.
Crooked functions can be used to construct many interesting
combinatorial objects, such as distance regular graphs (see
[1,27,28]).
\par A function $f:
F_{2^n}\rightarrow F_{2^n}$ is called quadratic if it is defined by
a polynomial with exponents of binary weight 2, i.e.,
\[
f(x)=\sum_{0\leq i,j\leq n-1}c_{ij}x^{2^i+2^j}.
\]
Clearly, $f(x)+f(y)+f(x+y)$ is bilinear, and therefore a quadratic
function is crooked if and only if it is APN. It has been proved
that the only crooked power functions are the quadratic functions
$x^{2^i+2^j}$ with $\gcd(n, i-j)=1$ [24,25], and a binomial function
$ax^i+bx^j$ can be crooked only if both exponents $i,j$ have binary
weight $\leq2$ [5].
\par If $f$ is an APN function, $A_1,A_2$ are affine permutations
and $A$ is an affine map, then the function $g=A_1\circ f\circ
A_2+A$ is also APN. The function $f$ and $g$ are then called
extended affine (EA) and simply affine equivalent if $A=0$. The
differentially uniformity of a function is an invariant of EA
equivalence. Besides, the inverse of any APN permutations is APN as
well. However, a permutation is not necessarily EA equivalent to its
inverse, even though they have the same differentially uniformity.
\par In [14], Carlet, Charpin and Zinoviev introduced a more general
notion of equivalence, referred to as Carlet-Charpin-Zinoviev (CCZ)
equivalent, which preserves APN and AB properties. Two functions $f$
and $g$ are called CCZ equivalent if for some affine permutation
${\Theta}$ of $F_{2^n}^2$, the image of the graph of $f$ is the
graph of $g$, i.e. ${\Theta}(G_f)=G_g$, where $G_f=\{(x,f(x)):x\in
F_{2^n}\}$ and $G_g=\{(x,g(x)):x\in F_{2^n}\}$. Differentially
uniformity and resistance to linear and differential cryptanalyses
are invariants of CCZ equivalence. Moreover, EA equivalence is a
particular case of CCZ equivalence and any permutation is always CCZ
equivalent to its inverse (see [23]).
\par It was believed that
any quadratic APN polynomial is affine equivalent to a Gold power
function ($x^{2^i+1}$ with $\gcd(n, i)=1$). In [20], A new APN
function on $F_{2^{10}}$ ($f(x)=x^3+ux^{36}$ for a suitable $u\in
F_{2^{10}}$) which is not affine equivalent to any of previous known
APN functions are constructed. In [8], Budaghyan and Carlet
construct a new infinite class of quadratic APN trinomials and a new
potentially infinite class of quadratic APN hexanomials which they
conjecture to be CCZ inequivalent to power functions for $n\geq 6$
and they confirm this conjecture for $n=6$. Then two new classes of
quadratic APN binomials CCZ inequivalent to power functions are
constructed in [10], those are the first found infinite classes of
APN polynomials which are proved not to be CCZ equivalent to power
function. Bierbrauer gives a brief construction in [3] for all known
examples of crooked binomials, which consist of an infinite family
and one sporadic example. In [7], the authors introduce two new
infinite classes of APN functions, one on fields of order $2^{2k}$
for $k$ not divisible by 2, and the other on fields of order
$2^{3k}$ for $k$ not divisible by 3. The polynomials in the first
class have between three and $k+2$ terms, and the second class's
polynomials have three terms.
\par In this paper, we introduce two infinite classes of quadratic crooked
multinomials on fields of order $2^{2m}$. The first class of APN
functions constructed in [7] is a particular case of the one we
construct in Theorem 1. Moreover, in Section 3, we prove that the
two classes of crooked multinomials constructed in this paper are
not EA equivalent to all power functions and we conjecture that CCZ
inequivalence between them also holds.

\section{Two classes of crooked multinomials}

\par To establish the crooked property of a function $f$ on
$F_{2^n}$, we must show that $f$ is APN, that is, the equation
$f(x)+f(x+a)=b$ has at most two solutions in $F_{2^n}$, for every
$a\neq 0$ and every $b$ in $F_{2^n}$. Moreover, we should show that
the set $\{f(x)+f(x+a): x\in F_{2^n}\}$ is an affine hyperplane of
$F_{2^n}$ for each $0\neq a\in F_{2^n}$. If $f$ is quadratic, then
$f$ is crooked if and only if $f$ is APN, and the equation has at
most two solutions if and only if $f(x)+f(x+a)+f(a)=0$ has two
solutions. In Theorem 1 and Theorem 2, we will prove that the two
infinite classes of quadratic multinomials constructed by us are APN, and therefore are crooked.\\
\par {Theorem 1.} Let $m,i,j$ be any positive integers such that $i>j$, and let $n=2m$,
$q=2^m$, $\gcd(i-j,n)=1$, $r_i\in F_{2^m}$ for each $i$, and $c,d\in
F_{2^n}$ be such that $c\notin F_{2^m}$, $d\notin \{u^{2^i+2^j},
u\in F_{2^n}\}$. Let $\{0,1\}\neq K\subseteq \{0,1,...,n-1\}$ be
such that $\sum_{k\in K}x^{2^k-1}$ is irreducible over $F_{2^n}$.
Then the multinomial
\begin{eqnarray*}
f(x) &=& cx^{q+1}+\sum_{i=1}^{m-1}r_ix^{2^i+q2^i} \\
&+& \sum_{k\in K}(d^{2^k}x^{2^{i+k}
+2^{j+k}}+d^{q2^k}x^{q(2^{i+k}+2^{j+k})})
\end{eqnarray*}
is crooked on $F_{2^n}$.
\par Proof. For any $0\neq a\in F_{2^n}$,
\begin{eqnarray*}
F(x) &=& f(x)+f(x+a)+f(a)\\
&=&
c(x^qa+xa^q)+\sum_{i=1}^{m-1}r_i(x^{2^i}a^{q2^i}+x^{q2^i}a^{2^i})\\
&+& \sum_{k\in K}d^{2^k}(x^{2^{i+k}}a^{2^{j+k}}+x^{2^{j+k}}a^{2^{i+k}})\\
&+& \sum_{k\in
K}d^{q2^k}(x^{q2^{i+k}}a^{q2^{j+k}}+x^{q2^{j+k}}a^{q2^{i+k}}).
\end{eqnarray*}
Since
\begin{eqnarray*}
F_1 &=& f(x)+f(x)^q=(c+c^q)x^{q+1},\\
F_2 &=& f(x+a)+f(x+a)^q=(c+c^q)(x+a)^{q+1},\\
F_3 &=& f(a)+f(a)^q=(c+c^q)a^{q+1},
\end{eqnarray*}
we have
\begin{eqnarray*}
F(x)+F(x)^q&=&F_1+F_2+F_3 \\
&=&(c+c^q)(x^qa+xa^q).
\end{eqnarray*}
If $F(x)=0$, then $F(x)+F(x)^q=0$. Since $c\notin F_{2^m}$, we have
$x^qa+xa^q=0$. Let $x=at$. Then $t^q=t$. The equation $F(x)=0$
becomes
\[
\sum_{k\in
K}((da^{2^{i}+2^{j}}+d^{q}a^{q(2^{i}+2^{j})})(t^{2^{i}}+t^{2^{j}}))^{2^k}=0.
\]
Since $\sum_{k\in K}x^{2^k-1}$ is irreducible over $F_{2^n}$ and it
is not equal to $x+1$, we get
\[
(da^{2^{i}+2^{j}}+d^{q}a^{q(2^{i}+2^{j})})(t^{2^{i}}+t^{2^{j}})=0.
\]
 $d\notin \{u^{2^i+2^j},
u\in F_{2^n}\}$ implies
\[
da^{2^{i}+2^{j}}+d^{q}a^{q(2^{i}+2^{j})}\neq 0,
\]
and therefore $t^{2^{i}}+t^{2^{j}}=0$, that is
$t^{2^{j}}(t^{2^{i-j}}+1)=0$. Since $\gcd(i-j,n)=1$, we get $t=0$ or
1. Therefore, $f(x)$ is APN, and the result follows.\\
Remarks: \\
1. The class of APN function constructed in Theorem 1 of [7] is a
particular case of the one we construct above for $j=0$, $K=\{0 \}$
and $m$ and $i$ both odd.\\
2. Let $m=6$, $i=8$, $j=1$, $K=\{0\}$, $c,d$ be primitive elements
of $F_{2^{12}}$ and $r_i\in F_{2^6}$ for each $i$. Then the function
\[
f(x) = cx^{2^6+1}+dx^{2^{8}
+2}+d^{2^6}x^{2^7+4}+\sum_{i=1}^{m-1}r_ix^{2^i+2^{i+6}}
\]
 is a crooked function of
$F_{2^{12}}$, which is an example not belong to the class
constructed in [7].\\
\par {Theorem 2.} Let $m,i,j$ be any positive integers such that $i>j$, and let $n=2m$,
$q=2^m$, $\gcd(i-j,n)=1$, and $c,d,r_i\in F_{2^n}$ be such that
$d^{q+1}=1$, $c+dc^q\neq 0$, $d\notin \{u^{2^i+2^j}, u\in F_{2^n}\}$
and $d=r_i^{1-q}$ for each $i$. Let $\{0,1\} \neq K\subseteq
\{0,1,...,n-1\}$ be such that $\sum_{k\in K}x^{2^k-1}$ is
irreducible over $F_{2^n}$. Then the multinomial
\begin{eqnarray*}
f(x) &=& cx^{q+1}+\sum_{i=1}^{m-1}r_ix^{2^i+q2^i} \\
&+& \sum_{k\in K}(x^{2^{i+k} +2^{j+k}}+dx^{q(2^{i+k}+2^{j+k})})
\end{eqnarray*}
is crooked on $F_{2^n}$.
\par Proof. For any $0\neq a\in F_{2^n}$,
\begin{eqnarray*}
F(x) &=& f(x)+f(x+a)+f(a)\\
&=&
c(x^qa+xa^q)+\sum_{i=1}^{m-1}r_i(x^{2^i}a^{q2^i}+x^{q2^i}a^{2^i})\\
&+& \sum_{k\in K}(x^{2^{i+k}}a^{2^{j+k}}+x^{2^{j+k}}a^{2^{i+k}})\\
&+& \sum_{k\in
K}d(x^{q2^{i+k}}a^{q2^{j+k}}+x^{q2^{j+k}}a^{q2^{i+k}}).
\end{eqnarray*}
we have
\[
F(x)+d\cdot F(x)^q=(c+dc^q)(x^qa+xa^q).
\]
If $F(x)=0$, then $F(x)+d\cdot F(x)^q=0$. Since $c+dc^q\neq 0$, we
have $x^qa+xa^q=0$. Let $x=at$. Then $t^q=t$. The equation $F(x)=0$
becomes
\[
\sum_{k\in
K}((a^{2^{i}+2^{j}}+da^{q(2^{i}+2^{j})})(t^{2^{i}}+t^{2^{j}}))^{2^k}=0.
\]
Since $\sum_{k\in K}x^{2^k-1}$ is irreducible over $F_{2^n}$ and it
is not equal to $x+1$, we get
\[
(a^{2^{i}+2^{j}}+da^{q(2^{i}+2^{j})})(t^{2^{i}}+t^{2^{j}})=0.
\]
 $d\notin \{u^{2^i+2^j},
u\in F_{2^n}\}$ implies
\[
a^{2^{i}+2^{j}}+da^{q(2^{i}+2^{j})}\neq 0,
\]
and therefore $t^{2^{i}}+t^{2^{j}}=0$, that is
$t^{2^{j}}(t^{2^{i-j}}+1)=0$. Since $\gcd(i-j,n)=1$, we get $t=0$ or
1. Therefore, $f(x)$ is APN, and the result follows.\\

\section{Their inequivalence with power crooked functions}

\par It is known
that the only crooked power functions on $F_{2^n}$ are the quadratic
functions $x^{2^i+2^j}$ with $\gcd(n, i-j)=1$, which are equivalent
to $x^{2^s+1}$ with $\gcd(n, s)=1$. Proving CCZ inequivalence of
functions is very difficult. In what follows, we prove that the
crooked functions introduced by us are not EA equivalent to all
power functions.
\par By Theorem 1, we have
\[
f(x)=cx^{2^6+1}+dx^{2^8+2}+d^{2^6}x^{2^7+4}
\]
is crooked on $F_{2^{12}}$, where $c,d\in F_{2^{12}}$ are
primitive.\\
\par {Theorem 3.} Let $c,d\in F_{2^{12}}$ are
primitive. Then the function
\[
f(x)=cx^{2^6+1}+dx^{2^8+2}+d^{2^6}x^{2^7+4}
\]
is EA inequivalent to power functions on $F_{2^{12}}$.
\par Proof. Suppose $f$ is EA equivalent to a power
function. Then $f$ is EA equivalent to $x^{2^s+1}$ for some nonzero
$s\in Z/12Z$. Hence there exist affine permutations $A_1(x),A_2(x)$
and an affine map $A(x)$ such that
\[
A_1\circ f=(A_1)^{2^s+1}+A.
\]
Let
\begin{eqnarray*}
A_1(x) &=& \sum_{i\in Z/12Z}a_ix^{2^i}\\
A_2(x)&=& \sum_{j\in Z/12Z}b_jx^{2^j}
\end{eqnarray*}
Then we have
\begin{eqnarray*}
 & & \sum_{i\in Z/12Z}a_i(cx^{2^6+1}+dx^{2^8+2}+d^{2^6}x^{2^7+4})^{2^i}\\
&=& \sum_{j,k\in Z/12Z}b_jb_k^{2^s}x^{2^j+2^{k+s}}+A
\end{eqnarray*}
Compare the coefficients of the terms with the same degree, we get
\begin{equation}
b_ib_{i+t-s}^{2^s}=b_{i+t}b_{i-s}^{2^s},
\end{equation}
for $t\in
Z/12Z$ and $t\neq 5,6,7$, and
\begin{equation}
a_ic^{2^i} = b_{i+6}b_{i-s}^{2^s}+b_ib_{i+6-s}^{2^s}
\end{equation}
\begin{equation}
a_id^{2^i} = b_{i+8}b_{i+1-s}^{2^s}+b_{i+1}b_{i+8-s}^{2^s}
\end{equation}
\begin{equation}
a_id^{2^{i+6}} = b_{i+7}b_{i+2-s}^{2^s}+b_{i+2}b_{i+7-s}^{2^s}
\end{equation}
Suppose $a_i\neq 0$ for some $i$. If $b_{i+j-s}\neq 0$, for some
$j$, then by (1) we have
\[
b_{i+j}b_{i+j-s}^{-2^s}=b_{i+j+t}b_{i+j+t-s}^{-2^s},
\]
for $t\in Z/12Z$ and $t\neq 5,6,7$.\\
If $j=0$, we
 take $t=1-j$ and $8-j$, then
\[
b_{i+j}b_{i+j-s}^{-2^s}=b_{i+1}b_{i+1-s}^{-2^s}=b_{i+8}b_{i+8-s}^{-2^s},
\]
which is contradictory to the equation (3). Therefore, we get
\begin{equation}
b_{i-s}=0
\end{equation}
 If $j=6$, we
 take $t=2-j$ and $7-j$, then
\[
b_{i+j}b_{i+j-s}^{-2^s}=b_{i+2}b_{i+2-s}^{-2^s}=b_{i+7}b_{i+7-s}^{-2^s},
\]
which is contradictory to the equation (4). Therefore, we get
\begin{equation}
b_{i+6-s}=0
\end{equation}
By (2), (5) and (6), we deduce that $a_i=0$. Then by the randomness
of $i$, we have $A_1(x)=0$, which is a contradiction, and the result
follows.
\par By the same method, we can deduce the following Theorem:\\
\par {Theorem 4.} The  crooked functions constructed in Theorem 1
and Theorem 2 are EA inequivalent to power functions on $F_{2^{n}}$.
\par We conjecture that the functions constructed in Theorem 1
and Theorem 2 are CCZ inequivalent to power functions, which we leave it as an open problem.\\
\par {Conjecture 1.} The  crooked functions constructed in Theorem 1
and Theorem 2 are CCZ inequivalent to power functions on
$F_{2^{n}}$.
\section{ Conclusion}
\par In this paper, we introduced two infinite classes of quadratic crooked
multinomials on fields of order $2^{2m}$. One class of APN
functions constructed in [7] is a particular case of the one we
constructed in Theorem 1. Moreover, we proved that the two classes
of crooked functions constructed in this paper are EA inequivalent
to power functions and conjectured that CCZ inequivalence between
them also holds.


\end{document}